\documentclass[acmsmall, screen, nonacm]{acmart}
\title{Decomposable Type Highlighting for Bidirectional Type and Cast Systems}
\author{Max Carroll}
\email{mc2372@cam.ac.uk}
\orcid{0009-0007-9636-8502}
\affiliation{%
  \institution{University of Cambridge}
  \department{Department of Computer Science and Technology}
  \city{Cambridge}
  \country{UK}
}
\author{Anil Madhavapeddy}
\email{avsm2@cam.ac.uk}
\orcid{0000-0001-8954-2428}
\affiliation{%
  \institution{University of Cambridge}
  \department{Department of Computer Science and Technology}
  \city{Cambridge}
  \country{UK}
}
\author{Patrick Ferris}
\email{pf341@cam.ac.uk}
\orcid{0000-0002-0778-8828}
\affiliation{%
  \institution{University of Cambridge}
  \department{Department of Computer Science and Technology}
  \city{Cambridge}
  \country{UK}
}
\date{June 2025}

\setcopyright{none}

\usepackage{subcaption}
\usepackage[dvipsnames]{xcolor}
\usepackage{semantic}  
\usepackage{listings}
\usepackage{stmaryrd}
\usepackage{wasysym}   

\newcommand{\scast}[2]{{\color{blue}\langle} #1 {\color{blue}\Rightarrow} #2 {\color{blue}\rangle}}
\newcommand{\scasterror}[2]{{\color{red}\langle} #1 {\color{red}\Rightarrow} \dyn {\color{red}\nRightarrow} #2 {\color{red}\rangle}}

\newcommand{\dyn}{{\color{teal}\texttt{?}}}

\newcommand{\synthesis}[3][\Gamma]{#1 \vdash #2 {\color{BrickRed}\ \Rightarrow}\ #3}
\newcommand{\analysis}[3][\Gamma]{#1 \vdash #2 {\color{BlueViolet}\ \Leftarrow}\ #3}

\newcommand{\hole}{\dyn}
\newcommand{\funmatch}{\blacktriangleright_\rightarrow}

\newcommand{\hlcmaths}[2][yellow!50]{\colorbox{#1}{$\displaystyle #2$}}

\newcommand{\gap}{{\color{teal}\Box}}
\newcommand{\type}[1]{\llbracket #1 \rrbracket}
\newcommand{\cmark}{{\color{blue}\Circle}}

\definecolor{commentgreen}{RGB}{2,112,10}
\definecolor{eminence}{RGB}{108,48,130}
\definecolor{weborange}{RGB}{255,165,0}
\definecolor{frenchplum}{RGB}{129,20,83}
\definecolor{purple}{RGB}{159,106,228}

\newcommand{\code}[1]{\text{\lstinline{#1}}}

\begin{document}

\begin{abstract}
We explore how to provide programmers with an interactive interface for explaining the process by which static types and dynamic casts are derived, with the goal of improving the debugging of static and dynamic type errors. To this end, we define mathematical foundations for a decomposable highlighting system within a bidirectional system and show how these can be propagated through dynamic types in a cast system.
Our prototype implementation in the gradually typed Hazel language includes a web-based user interface, through which we highlight the importance of type-level debugging.
\end{abstract}

\maketitle

\section{Introduction}

In static typing, blame for type errors is typically localised to a single location in the code. However, this localisation may be misleading, as the actual cause of the error might be rooted in a broader context. For example, in OCaml 65\% of type errors are related to \textit{multiple} locations~\cite{StudentTypeErrorFixes} and furthermore, the errors only state the expected types without explanation for \textit{why} they occur.
In dynamic typing, errors do not typically specify any source code context that caused them, instead relying on the interpretation of (potentially complex and extensive) execution traces.

\paragraph{Vision:} We seek to improve user understanding of static and dynamic type systems and type errors by providing a \textit{formally} complete and decomposable highlighting system for bidirectional type systems (type slicing) and propagation of this information through dynamic cast systems (cast slicing). This would allow users to interactively explore \textit{why} an expression has been typed by decomposing the highlighted segments by their influence on the expression's type, inspecting only the particular parts they do not understand. Further, in languages with dynamic type information, this highlighting information can be propagated through, for example, dynamic casts, providing source code context to explain why a cast was executed during evaluation.\footnote{For either a compiler-generated or user-inserted cast.}

\paragraph{Progress:} This paper lays out our approach to building mathematical foundations using Hazel \cite{Hazel}, a research language that allows incomplete programs (with holes) with a focus on liveness, interaction design \cite{Grove, StructureEditing, LiveLits} and learning ~\cite{HazelTutor, HazelProofAssistant}. Hazel is gradually typed, so these highlighting methods can be explored for explaining both static and dynamic types (and type errors). However, the foundations of this work apply more generally to many bidirectional type systems and cast systems.
While this paper focuses on giving definitions and expected properties\footnote{No proof mechanisation has been performed as of yet. But `informal' proofs of most properties have been explored.} of a formal foundation, a preliminary implementation is also deployed at \url{https://hazel.org/build/witnesses-type-slicing/} for example usage.

\section{Background}
First, we introduce the notions of bidirectional types, cast systems, gradual types, and the core Hazel calculus for reference.

\subsection{Bidirectional Type Systems}\label{sec:BidirectionalTypeSystem}
A \textit{bidirectional type system} \cite{BidirectionalTypes} takes on a more algorithmic definition of typing judgements, being more intuitive to implement, while still allowing some amount of local type inference.

This is done in a similar way to annotating logic programs \cite[pg. 123]{LogicProg}, by specifying the \textit{mode} of the type parameter in a typing judgement, distinguishing when it is an \textit{input} (type analysis/checking) and when it is an \textit{output} (type synthesis). We express this with two judgements, read respectively as: \textit{$e$ synthesises an (\text{output}) type $\tau$/analyses against an (\text{input}) type $\tau$ under typing context $\Gamma$}:
\[\synthesis{e}{\tau} \qquad\qquad \analysis{e}{\tau}\]


Such languages should be \textit{mode correct}\footnote{Ensuring that they can be easily implemented algorithmically. That is, never require the `guessing' of inputs.} \cite{ModeCorrectness} and will have three obvious rules. That variables can synthesise their type, if it is accessible from the typing assumptions. Annotated terms synthesise their type from the annotation. Subsumption: a synthesising term must check against that same type.
\begin{figure}
\begin{subfigure}{0.28\textwidth}
    \[\inference{x : \tau \in \Gamma}{\synthesis{x}{\tau}}\]
    \caption{Variables synthesise their type from the typing assumptions.}
\end{subfigure}
$\qquad$
\begin{subfigure}{0.28\textwidth}
    \[\inference{\analysis{e}{\tau}}{\synthesis{e : \tau}{\tau}}\]
    \caption{Annotations synthesise the annotated type.}
\end{subfigure}
$\qquad$
\begin{subfigure}{0.28\textwidth}
    \[\inference{\synthesis{e}{\tau}}{\analysis{e}{\tau}}\]
    \caption{Subsumption: A synthesising term checks against the same type.}
\end{subfigure}
\end{figure}

\subsection{(Dynamic) Cast Calculi}\label{sec:DynamicTypeSystem}
A cast calculus adds casts between types to an operational semantics. More specifically, we consider a dynamically typed system, with a distinguished dynamic type, notated \dyn.

Cast expressions will be denoted $e\scast{\tau_1}{\tau_2}$ for expression $e$ and types $\tau_1, \tau_2$, representing that $e$ has type $\tau_1$ and is cast to new type $\tau_2$. Compound type casts can be decomposed during evaluation. For example, applying $v$ to a function wrapped in a cast decomposes the cast into casting the applied argument and then the result:
\[(f\scast{\tau_1 \to \tau_2}{\tau_1' \to \tau_2'})(v) \mapsto (f(v \scast{\tau_1'}{\tau_1})\scast{\tau_2}{\tau_2'})\]
Or if $f$ has the dynamic type, it should still be treated as a possible function:
\[(f\scast{\dyn}{\tau_1' \to \tau_2'})(v) \mapsto (f(v \scast{\tau_1'}{\dyn})\scast{\dyn}{\tau_2'})\]

Hence, casts around functions (type information) will be moved to the actual arguments at runtime, meeting with casts casts on the argument, resulting in a cast error or a successful cast to a corresponding value of the new type. The cast on the argument is reversed, in a similar vein to the contravariance of function argument types under sub-typing.

\subsection{Gradual Type Systems}\label{sec:GradualTypeSystem}

A \textit{gradual type system} \cite{GradualRefined, GradualFunctional} combines static and dynamic typing. Terms may be annotated as dynamic, marking regions of code `omitted' from type-checking but still \textit{interoperable} with static code. For example, the following (pseudo-OCaml syntax) type checks:
\begin{lstlisting}[escapeinside=||]{reason}
let x : |\dyn| = 10 in  /* Dynamically typed */
x ^ "str"          /* Statically typed */
\end{lstlisting}
Where \code{^} is string concatenation expecting inputs to be \code{string}. But would then cause a runtime \textit{cast error} when attempting to calculate \code{10 ^ "str"}. Typically, the language is split into two parts:
\begin{itemize}
\item The \textit{external language} -- where static type checking is performed which allows annotating expressions with the dynamic type.
\item The \textit{internal language} -- where evaluation and runtime type checking is performed via cast expressions.\footnote{i.e. the proposed \textit{dynamic type system} above.} The example above would reduce to a \textit{cast error}: \[\code{10}\scasterror{\code{Int}}{\code{String}}\ \code{^ "str"}\]
\end{itemize}
This is possible by introduction of a \textit{consistency} equivalence relation notated $\tau_1 \sim \tau_2$ used in place of type equality in the typing rules. Consistency is a weakening of equality: all types are consistent with the dynamic type $\dyn$, and compound types are consistent if their sub-parts are:
\[\inference{}{\tau \sim \dyn}\qquad \inference{\tau_1 \sim \tau_1' & \tau_2 \sim \tau_2}{\tau_1 \to \tau_2 \sim \tau_1' \to \tau_2'}\]

Finally, the static type information needs to be encoded into casts to be used in the dynamic internal language, for which the evaluation semantics are defined. This is done by \textit{elaboration}, $\Gamma \vdash e \leadsto d : \tau$ read as: \textit{external expression $e$ is elaborated to internal expression $d$ with type $\tau$ under typing context $\Gamma$}. For example, to insert casts around function applications:
\[\inference{\Gamma \vdash e_1 \leadsto d_1 : \tau_1 & \Gamma \vdash e_2 \leadsto d_2 : \tau_2' \\ \tau_1 \funmatch \tau_2 \to \tau  & \tau_2 \sim \tau_2'}{\Gamma \vdash e_1(e_2) : \tau \leadsto (d_1\scast{\tau_1}{\tau_2 \to \tau})(d_2\scast{\tau_2'}{\tau_2}) : \tau}\]
Where $\funmatch$ explicitly pattern matches function types, including $\dyn$ where $\dyn \funmatch \dyn \to \dyn$. We place a cast on the function\footnote{This cast is required, as if $\tau_1 = \dyn$ then we need a cast to realise that it is even a function. Otherwise $\tau_1 = \tau_2 \to \tau$ and the cast is redundant.} $d_1$ to $\tau_2 \to \tau$ and on the argument $d_2$ to the function's expected argument type $\tau_2$ to perform runtime type checking of arguments.
Intuitively, casts must be inserted whenever type consistency is used, but deciding which casts to insert is non-trivial \cite{Gradualizer}.

The runtime semantics of the internal expression is that of the \textit{dynamic cast system} discussed above (\ref{sec:DynamicTypeSystem}). A cast is succeeds if and only if the types are \textit{consistent}.

\subsection{The Hazel Calculus}\label{sec:CoreHazel}
\index{Hazel}
Hazel is built upon a bidirectional and gradually typed core lambda calculus \cite{HazelLivePaper}. It additionally includes \textit{holes}, which can both be typed and treated as an indeterminate \textit{final form} (value), allowing evaluation to proceed around them seamlessly. Errors can be treated as holes (i.e. with dynamic/unknown type) to allow for continued evaluation.

The core calculus \cite{HazelLivePaper} is a gradually and bidirectionally typed lambda calculus. Therefore it has a gradual and locally inferred bidirectional \textit{external language} elaborated to an explicitly typed \textit{internal language} including cast expressions. Holes will also be notated by $\hole$ and naturally synthesise the dynamic type.\footnote{Notation here differs from the original Hazel paper \cite{HazelLivePaper}, in order to be consistent with Hazel UI. Hole meta-variables and non-empty holes are of little interest to this paper, so are omitted.}

\section{Preliminary Implementation}
\begin{figure}[ht]
\centering
\begin{subfigure}[t]{0.45\textwidth}
\centering
\includegraphics[width=1\textwidth]{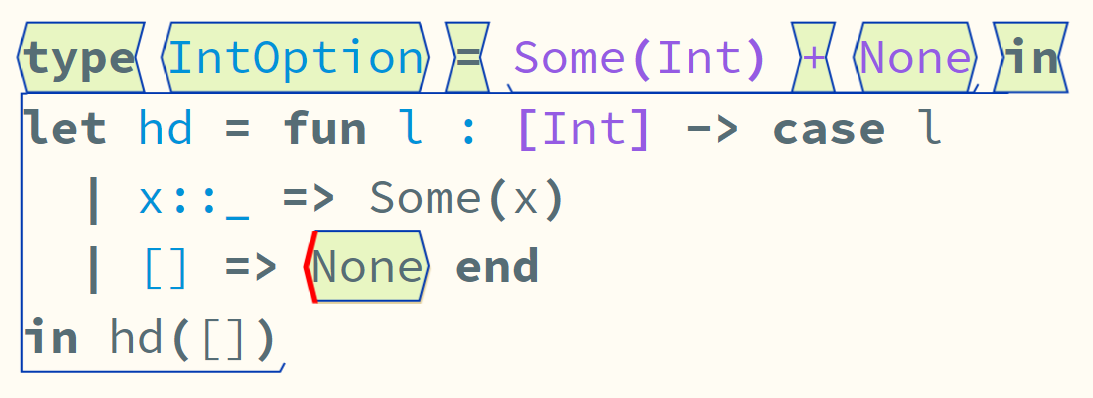}
\caption{\code{None} synthesises \code{IntOption} due to it being a value of \code{IntOption}. The assumption that \code{None} has type \code{IntOption} results in the slice of its definition to also be highlighted, notice also the inclusion of the alias binding for \code{IntOption}}.
\end{subfigure}$\qquad$
\begin{subfigure}[t]{0.45\textwidth}
\centering
\includegraphics[width=1\textwidth]{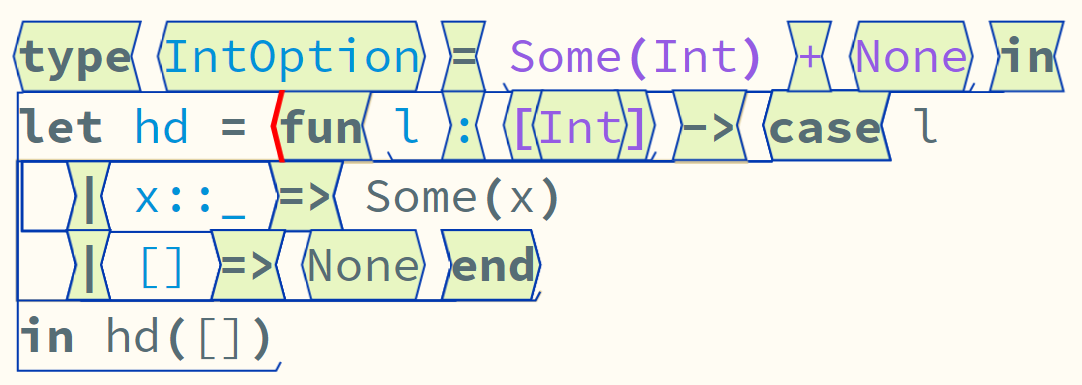}
\caption{The function synthesises \code{[Int]}$\to$ \code{IntOption} due to its \code{[Int]} annotation and that the match branches synthesis \code{IntOption}. Both branches provide the same type information, so only one branch (the last) is highlighted.}
\end{subfigure}
\begin{subfigure}[t]{0.45\textwidth}
\centering
\includegraphics[width=1\textwidth]{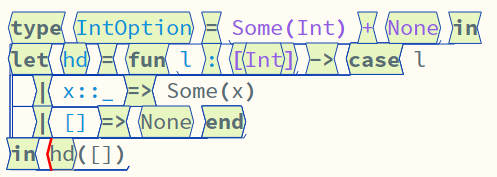}
\caption{The variable \code{hd} synthesises \code{[Int]}$\to$ \code{IntOption} by assumption similarly to (a). The slice of the definition of \code{hd} is also highlighted.}
\end{subfigure}$\qquad$
\begin{subfigure}[t]{0.45\textwidth}
\centering
\includegraphics[width=1\textwidth]{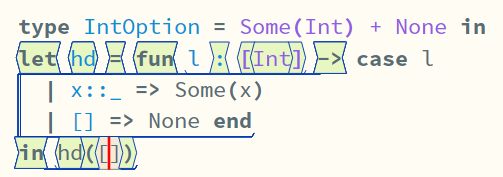}
\caption{The list input is expected to be an \code{[Int]} as it is applied to \code{hd} which is a function annotated with input type \code{[Int]}.}
\end{subfigure}
\caption{Type Slicing Examples}
\label{fig:SliceExamples}
\end{figure}

Next, we briefly demonstrate, by example in Hazel, the ideas of type and cast slicing, before diving into the mathematical foundations. Figure \ref{fig:SliceExamples} shows the type slices of four sub-expressions at the cursor (in red). Typing of a sub-expression is done under typing assumptions; the UI also highlights the type slice of a variable or type definition if the assumption is required in order to type check.\footnote{For example, the \code{IntOption} type definition, or the \code{hd} binding, in fig. \ref{fig:SliceExamples}}

\begin{figure}[ht]
\centering
\begin{subfigure}{0.45\textwidth}
\centering
\includegraphics[width=1\textwidth]{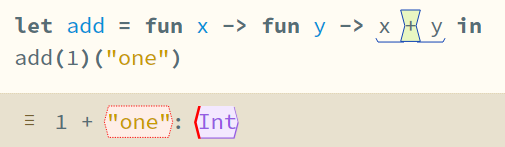}
\caption{A simple cast error blaming the plus operator for requiring the integer cast.}
\end{subfigure}$\qquad$
\begin{subfigure}{0.45\textwidth}
\centering
\includegraphics[width=1\textwidth]{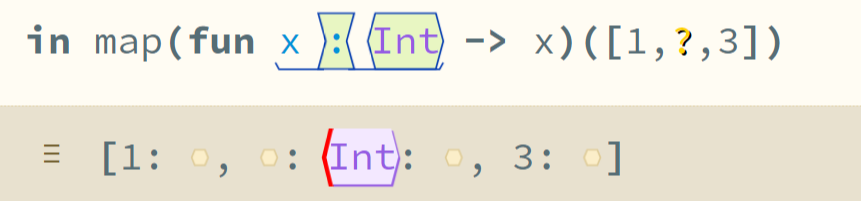}
\caption{A hole (notated `?') cast to \code{Int} due to being the argument of a mapped function annotated with an \code{Int} input.}
\end{subfigure}
\caption{Cast Slicing Examples}
\label{fig:CastSliceExamples}
\end{figure}

Casts between types are inserted around expression, or explicitly by the programmer. Type slices can be associated with these casted types, highlighting the source code that enforced their automated insertion by the compiler during elaboration, or the corresponding cast written in the code directly by the programmer. Figure \ref{fig:CastSliceExamples} demonstrates two simple examples, the first of which shows how cast slicing could be use to perform error highlighting for dynamic errors.\footnote{Both these examples select the type being casted to. But, the system works equally well selecting the castee type instead; however, the Hazel UI does not yet have a way to select the castee type.}

\section{Type Slicing Theory}\label{sec:TypeSlicingTheory}
This section details the underlying mathematical foundation, first defining some preliminary constructs, used to define two core slicing criteria: \textit{Synthesis Slices}, and \textit{Analysis Slices}, which minimally highlight the parts (slice) of a term directly causing the term to \textit{synthesise} a type or of its context\footnote{The code surrounding the term.} enforcing it to \textit{analyse against} some type.

\subsection{Expression Typing Slices}\label{sec:ExpressionTypingSlices}
First, we introduce what \textit{slices} are. The aim is to provide a formal representation of term highlighting. 

\subsubsection{Term Slices}
A \textit{term slice} is a term with some sub-terms omitted. The omitted terms are those that are \textit{not} highlighted. For example, if my slicing criterion is to \textit{omit terms which are typed as} \code{Int}, then the following expression highlights as shown on the left. This is encoded, on the right, by representing omitted sub-terms by \textit{gap} terms, notated $\gap$:

\[\hlcmaths[yellow!30]{(\lambda} x: \code{Int}\hlcmaths[yellow!30]{.\ \lambda y : \code{Bool}.}\ x\hlcmaths[yellow!30]{)(}1\hlcmaths[yellow!30]{)} \qquad\qquad (\lambda \gap.\ \lambda y : \code{Bool}.\ \gap)(\gap)\]

We can then define a \textit{precision} partial order \cite{PartialOrder} on term slices: $\varsigma_1 \sqsubseteq \varsigma_2$ meaning $\varsigma_1$ is less or equally precise than $\varsigma_2$. That is, $\varsigma_1$ matches $\varsigma_2$ structurally except that some sub-terms may be gaps. For example:
\[\gap \sqsubseteq\gap + \gap\sqsubseteq 1 + \gap \sqsubseteq 1 + 2\]

\paragraph{Lattice Structure:}\label{sec:JoinTypesTheory} For any \textit{complete term} $t$ (having no gaps), the slices of $t$ form a \textit{bounded lattice structure} \cite{Lattice}. That is, every pair $\varsigma_1, \varsigma_2$ has a \textit{join} $\varsigma_1 \sqcup \varsigma_2$ and \textit{meet} $\varsigma_1 \sqcap \varsigma_2$. In general, not all slices have joins: $1 \centernot\sqcup 2$, but do have meets as $\gap \sqsubseteq \varsigma$ for all $\varsigma$.
 
\subsubsection{Typing Assumption Slices}
Expression typing is performed given a set of \textit{typing assumptions}. Therefore, in addition, we also desire a slice taking the \textit{relevant} assumptions. We represent typing assumptions by \textit{partial functions} mapping variables to types. Hence, their slices are just partial functions to \textit{type slices}. A slice must map a (possibly equal) subset of the variables to less or equally precise types. Precision, meets, and joins, can be defined pointwise:

\begin{definition}[Typing Assumption Slice Precision]
For typing assumption slices $\gamma_1, \gamma_2$. Where $\mathrm{dom}(f)$ is the set of variables for which a partial function $f$ is \textit{defined}:
\[\gamma_1 \sqsubseteq \gamma_2 \iff \mathrm{dom}(\gamma_1) \subseteq \mathrm{dom}(\gamma_2) \text{ and } \forall x \in  \mathrm{dom}(\gamma_1).\ \gamma_1(x) \sqsubseteq \gamma_2(x)\]
\end{definition}
\begin{definition}[Typing Assumption Slice Joins and Meets]
For typing slices $\gamma_1, \gamma_2$, and any variable $x$: 
\begin{itemize}
\item If $\gamma_1(x) = \bot$ then $(\gamma_1 \sqcup \gamma_2)(x) = \gamma_2(x)$ and $(\gamma_1 \sqcap \gamma_2)(x) = \bot$. 
\item Analogously if $\gamma_2(x) = \bot$.
\item Otherwise, $(\gamma_1 \sqcup \gamma_2)(x) = \gamma_1(x) \sqcup \gamma_2(x)$.
\end{itemize}
\end{definition}
Again, slicing complete typing assumptions $\Gamma$ forms a bounded lattice. In general, some slices have no join: consider $x : \code{Int}$ and $x : \code{String}$.

\subsubsection{Expression Typing Slices}
An \textit{expression typing slice}, $\rho$, is a pair, $\varsigma^\gamma$, of a term slice and a typing slice. Precision, joins and meets, can be extended componentwise to term typing slices with all the same properties. These slices are the core construct for synthesis slices.

\subsubsection{Typing} \textit{Expression slices} can be \textit{type checked} under the \textit{type assumption slices} by replacing gaps $\gap$ by: holes of any meta-variable $\hole$ in \textit{expressions}, fresh variables in \textit{patterns}, and the dynamic type in \textit{types} (notated by $\type{-}$). Other (non-gradual) systems require different interpretations, for example, a value of polymorphic type $\forall \alpha. \alpha $ could be used in place of gaps. Some form of polymorphism is required in order to determine at what point we have removed so much information that the sliced term has a \textit{more general} (more polymorphic) type than before.

\begin{definition}[Expression Typing Slice Type Checking]
For expression typing slice $\varsigma^{\gamma}$ and type $\tau$. $\synthesis[\gamma]{\varsigma}{\tau}$ iff $\synthesis[\type{\gamma}]{\type{\varsigma}}{\tau}$ and $\analysis[\gamma]{\varsigma}{\tau}$ iff $\analysis[\type{\gamma}]{\type{\varsigma}}{\tau}$.
\end{definition}
\subsection{Context Typing Slices}\label{sec:ContextTypingSlices}
An expression's analysing type is enforced by its \textit{surrounding context}. We must note a clash in terminology between \textit{contexts} of a term (the part of a program surrounding a sub-term, such as those as used in contextual dynamics) opposing the \textit{typing contexts} used to refer to typing assumptions made during type checking in the standard literature. For clarity, we refer to the typing assumptions as directly as is, and never refer to them as `contexts'.

For example, the  type of the underlined expression below is enforced by the surrounding highlighted context (the annotation):
\[\underline{(\lambda x. \hole )} \hlcmaths[yellow!30]{:  \code{Bool} \hlcmaths[yellow!30]{\to \code{Int}}}\]

\subsubsection{Contexts and Their Slices}
\newcommand{\C}{\mathcal{C}}
We represent these surrounding contexts by a \textit{term context} $\C$, which marks \textit{exactly one} sub-term as $\cmark$. Where $\C\{t\}$ substitutes term $t$ for the mark $\cmark$ in $\C$\footnote{Only allowed if the marked position expects a term of the same class as $t$ (pattern \code{Pat}, type \code{Typ}, or expression \code{Exp}).} and composition is defined as substituting contexts into contexts, notated infix by $\circ$.

\newcommand{\Cs}{c}
\newcommand{\p}{d}
Contexts extend to context slices analogously to term slices and are notated as $\Cs$. However, the precision relation $\sqsubseteq$ is more restrictive, requiring the mark $\cmark$ to remain in the same structural position. For example: $\cmark(\gap) \sqsubseteq \cmark(1)$, but $\cmark \not \sqsubseteq \cmark(1)$. This can be concisely defined pointwise.


Joins and meets can be defined pointwise as before, still forming bounded lattices over complete contexts. The lattice bottom is the \textit{purely structural context}, consisting of only gaps with the mark in the correct position. In general, in addition to joins, not all contexts have meets: $\cmark\centernot \sqcap \cmark(\gap)$.

\subsubsection{Typing Assumption Contexts and Their Slices}
The accompanying typing notion can be represented by \textit{endomorphisms on typing assumption slices}. These functions represents which \textit{relevant} typing assumptions must be \textit{added}, and those safely \textit{removable} when typing an expression within a context slice.

\newcommand{\F}{\mathcal{F}}
\newcommand{\f}{f}

Precision, joins, and meets can be defined pointwise, forming bounded lattices on complete functions as usual. The bottom element being the constant function to the empty typing assumptions. 

\subsubsection{Context Typing Slices}
An \textit{expression context typing slice}, $\p$, is a context slice with each sub-context recursively tagged by typing assumption context slices. Then, retrieve the underlying context with $\text{ctx}(\p)$ and typing assumption context with $\text{typ}(\p)$ \footnote{By composing all individual contexts up from the mark $\cmark$}. As before, lattice relations are defined componentwise. Gaps can be interpreted by holes during type checking analogously to expression typing slices.

\subsection{Indexed Type Slices}\label{sec:TypeIndexedSlices}
Decomposing slices of expressions with compound types (i.e. functions) according to their component types is the core idea of this method. For example, the following context slice on the left explains why the underlined term analyses $\code{Bool} \to \code{Int}$, and the right is a sub-slice explaining only the argument type:
\[\underline{(\lambda x. \hole )} \hlcmaths[yellow!30]{:  \code{Bool} \hlcmaths[yellow!30]{\to \code{Int}}} \qquad \underline{(\lambda x. \hole )} \hlcmaths[yellow!30]{:}  \code{Bool} \hlcmaths[yellow!30]{\to \code{Int}}\]
This has many uses which are core to both the user experience and the internal implementation: 
\begin{itemize}
    \item The programmer could use this to query sub-slices for only the sub-parts of the type of an expression that they do not already understand, hence omitting unneeded information and reducing the amount of highlighting. For example, if they knew why the argument of a function was an integer, but not why the return type was.
    \item Casts are decomposed throughout evaluation: a cast between function types will be separated into a cast between the two argument types and a cast between the two return types. Being able to decompose slices allows only the relevant source code to be retained.
    \item Internally, calculating slices is easier when sub-slices are available, during function application. See fig. \ref{fig:AnalysisSliceApplication}.
\end{itemize}
\newcommand{\Sb}{\mathcal{S}}
\newcommand{\s}{s}
The main property that indexed-slices should maintain is \textit{reconstructability}: that slices can be reconstructed from their sub-parts by joining the sub-slices. As sub-slices may slice different regions of code, we pair them with contexts which place the sub-slices within the same context, making them join-able. We only consider \textit{context slices} here, but expressions slices are type-indexed analogously. Context slices are syntactically defined to correspond with the structure of types:
\[\Sb ::= \p \mid \p * \Sb \to \p * \Sb\] 
But they must be \textit{reconstructable}. That is, the \textit{full slice} of $\Sb$, notated $\overline{\Sb}$ must exist by joining the sub-slices within their contexts:
\[\overline{\p} = \p \qquad \qquad \overline{\p_1 * \Sb_1 \to \p_2 * \Sb_2} = \p_1 \circ \overline{\Sb_1} \sqcup \p_2 \circ \overline{\Sb_2}\]

\subsection{Synthesis Slices}
\label{sec:SynthesisSlices}

Synthesis slices aim to explain why an expression \textit{synthesises} a type. They omit all sub-terms which analyse against a type retrieved from synthesising some other part of the program. For example, the following term synthesises a $\code{Bool} \to \code{Bool}$ type, and the variable $x : \code{Int}$ and argument are irrelevant:

\[\hlcmaths[yellow!30]{(\lambda} x: \code{Int}\hlcmaths[yellow!30]{.\ \lambda y : \code{Bool}.\ y)(}1\hlcmaths[yellow!30]{)}\]

\begin{definition}[Synthesis Slices]
For a synthesising expression, $\synthesis{e}{\tau}$. A synthesis slice is an expression typing slice $\varsigma^{\gamma}$ of $e^\Gamma$ which also synthesises $\tau$, that is, $\synthesis[\type{\gamma}]{\type{\varsigma}}{\tau}$.
\end{definition}
\begin{proposition}
A minimum synthesis slice of $\synthesis{e}{\tau}$, under $\sqsubseteq$, exists.
\end{proposition}

\subsection{Analysis Slices}\label{sec:AnalysisSlices}
A similar idea can be devised for type analysis, represented using \textit{context slices}. After all, it is the terms immediately \textit{around} the sub-term where the type checking is enforced. For example, when checking this annotated term on the left, the \textit{inner hole term} $\hole$ (underlined) must be consistent with \code{Int} due to the annotation and lambda constructor within its context, giving:
\[(\lambda x. \underline{\hole}) : \code{Bool} \to \code{Int} \qquad \hlcmaths[yellow!30]{(\lambda} x\hlcmaths[yellow!30]{.} \underline{\hole} \hlcmaths[yellow!30]{) : } \code{Bool} \hlcmaths[yellow!30]{\to \code{Int}}\]

In other words, if the inner hole was type checked within the context slice, then it would \textit{still} be required to analyse against \code{Int}. However, the overall synthesised type of the whole context may differ: the above would synthesise $\dyn \to \code{Int}$ vs. the original $\code{Bool} \to \code{Int}$.

\subsubsection{Checking Context}
\label{sec:CheckingContexts}
We only want to consider the smallest context \textit{scope} that enforced the type checking. For example, the below term has 3 annotations, but only the inner one enforces the \code{Int} type on the integer 1. I refer to this as the \textit{minimally scoped checking context}:
\[\underline{1} \hlcmaths[yellow!30]{: \code{Int}} : \dyn : \code{Bool}\]

\begin{definition}[Checking Context]
\label{def:CheckingContext}
If $\analysis[\Gamma]{e}{\tau}$. Then, a checking context for $e$ is a typing context $\p$ such that: $\text{ctx}(\p) \neq \cmark$, and $\synthesis[\text{typ}(\p)(\Gamma)]{\text{ctx}(\p)\{e\}}{\tau'}$ for some $\tau'$ while still retaining the sub-derivation for $\analysis[\Gamma]{e}{\tau}$.
\end{definition}
\begin{definition}[Minimally Scoped Checking Context]
For a derivation $\analysis[\Gamma]{e}{\tau}$, a minimally scoped expression checking context is a checking context of $e$ such that no sub-context is also a checking context.
\end{definition}
Observant readers will notice that any expression has infinitely many checking contexts. But importantly, there are only finitely many checking contexts and \textit{exactly one} minimally scoped checking context for a sub-expression which is itself (after substituting the sub-expression into its checking context) a sub-expression of a particular program.

\begin{figure}[h]
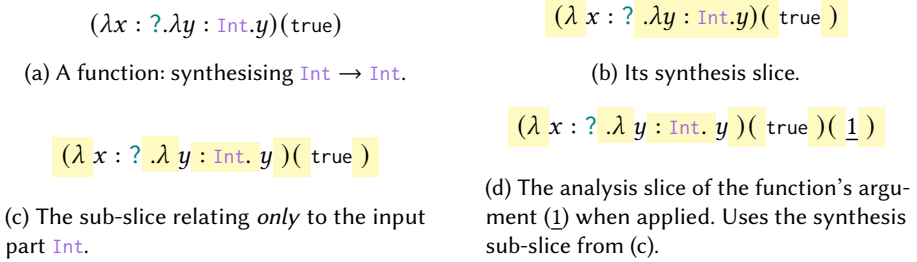

\centering
\begin{subfigure}{0.4\textwidth}
\[(\lambda x : \dyn. \lambda y : \code{Int}. y)(\code{true})\]
\caption{A function: synthesising $\code{Int} \to \code{Int}$.}
\end{subfigure}$\qquad$
\begin{subfigure}{0.4\textwidth}
\[\hlcmaths[yellow!30]{(\lambda} x : \dyn\hlcmaths[yellow!30]{. \lambda y : \code{Int}. y)(}\code{true}\hlcmaths[yellow!30]{)}\]
\caption{Its synthesis slice.}
\end{subfigure}
\begin{subfigure}{0.4\textwidth}
\[\hlcmaths[yellow!30]{(\lambda} x : \dyn\hlcmaths[yellow!30]{. \lambda} y \hlcmaths[yellow!30]{: \code{Int}.} y\hlcmaths[yellow!30]{)(}\code{true}\hlcmaths[yellow!30]{)}\]
\caption{The sub-slice relating \textit{only} to the input part $\code{Int}$.}
\end{subfigure}$\qquad$
\begin{subfigure}{0.4\textwidth}
\[\hlcmaths[yellow!30]{(\lambda} x : \dyn\hlcmaths[yellow!30]{. \lambda} y \hlcmaths[yellow!30]{: \code{Int}.} y\hlcmaths[yellow!30]{)(}\code{true}\hlcmaths[yellow!30]{)(}\underline{1}\hlcmaths[yellow!30]{)}\]
\caption{The analysis slice of the function's argument ($\underline{1}$) when applied. Uses the synthesis sub-slice from {(c)}.}
\end{subfigure}
\caption{Demonstration: Analysis slice application uses synthesis slices}
\label{fig:AnalysisSliceApplication}
\end{figure}

\begin{definition}[Analysis Slice]\label{def:analysisslice}
For $\analysis{e}{\tau}$ with a minimally scoped checking context $\p$. An analysis slice is a context slice $\p'$ of $\p$ where $\type{\p'}$ is also a checking context for $e$.
\end{definition}
\begin{proposition}\label{conj:AnalysisSliceUniqueness}
A minimum analysis slice of $\analysis{e}{\tau}$ in a checking context $\p$, under $\sqsubseteq$, exists.
\end{proposition}
Fig. \ref{fig:AnalysisSliceApplication} demonstrates an example of how this works for a more complex situation where function application enforces a type upon its argument.

\section{Cast Slicing Theory}\label{sec:CastSlicingTheory}
Cast slicing propagates type slice information during evaluation, by tagging casts types with type slices. The first two criteria work together during elaboration, inserted depending on if the cast component was derived from either type synthesis or analysis. The current rules are very involved, but build directly upon the Hazel calculus elaboration semantics. Future work will aim to simplify these rules.

\paragraph{Comparison with Blame:}
The idea of tagging information to casts is reminiscent of blame \cite{Blame} in gradual typing. Blame determines whether a cast error is caused by the expression within a cast or the context around a cast, always blaming more dynamic code. Parallels between synthesis (of the expression) and analysis (of the context) slices can be seen, but these are actually orthogonal to blame, with type slicing concerning why casts are \textit{inserted}. Future work could explore these connections and the integration of blame, with applications for type error debugging: we could point to exactly which portion of dynamic code (e.g. one particular static error inserted into a hole) is responsible for the dynamic error.

\section{The Hazel Implementation}
The implementation is under active development, can be found on the \texttt{witnesses-type-slicing} branch available at GitHub \cite{HazelCode}. Hazel extends the core calculus with many advanced features including:\footnote{As of July 2025} Lists, Tuples, Labelled Tuples (records) \cite[ch. 11.7-8]{TAPL}, Sum Types \cite[ch. 11.10]{TAPL}, Type Aliases, Pattern Matching, Explicit System F Style Polymorphism \cite[ch. 23]{TAPL}, Iso-Recursive Types \cite[ch. 22-23]{TAPL}. Type and cast slicing extends to these relatively simply, but, polymorphism and recursive types will require further (less trivial) extensions to the meta-theory.

\section{Future Work}
Future work aims to build upon these mathematical foundations in order to improve and explore the effectiveness of the human aspects of these highlighting systems.
\paragraph{Exploiting Decomposability in the UI:}
The merits of this formal system stem from the ability to deconstruct slices by their type. Uses include refining the highlighting to explain exactly which code corresponds to a specific subpart of the expression's type. For example, a user may understand why an expression is a function, but not why it has a particular return type; conversely, they may only not understand why the expression is a function, but not care about the argument or return types themselves. We wish to further explore how to use this information to provide a more \textit{intuitive and interactive UI} for use in the Hazel editor. This may include further refinements such as hiding, summarising, and jumping to the slices of variable definitions in an expression's type assumptions slice.
\paragraph{Polymorphism and Recursive Types:}
As previously mentioned, extending the meta-theory and implementation to parametric polymorphic systems and recursive types is planned. 
\paragraph{Decomposable Type Eliminators:}
The slice of a function application will include the application and part of the slice of the function, all compressed within a single type constructor, and therefore not decomposable (see sub-figure (b) in figure \ref{fig:AnalysisSliceApplication}).
In the future, we wish to improve this situation, potentially by creating a direct correspondence between derivations and slices, allowing indexing on both derivations \textit{and} types. Then slices could be decomposable according to the typing rules, for which an interactive UI could highlight code for each typing rule and even show/explain the rules in a pop-up similarly to the Explain This framework in Hazel \cite{HazelTutor}. This would require an entirely new, or at least major extension, to the presented theories in this paper.

Other ideas to deal with this which are purely to do with the UI, not requiring new formalisation, include emphasising\footnote{i.e. in a darker shade} the fundamental sub-expression that 'sourced' the type (inside of the type eliminator), For example, emphasising the inner lambda in fig. \ref{fig:AnalysisSliceApplication} (b).

\paragraph{Cast Slicing:}
Understanding how and why a cast was manipulated throughout evaluation requires inspecting potentially long and complex evaluation traces. UI to simplify evaluation traces to focus only on specific casts would be of use here. Additionally, there is scope to develop \textit{dynamic} slicing methods which would highlight minimal programs that evaluate to (a possibly less precise value) involving the same cast. These could be more akin to dynamic slicing in imperative languages \cite{DynProgSlice}, or in functional languages \cite{FunctionalProgExplain}. As Hazel can evaluate incomplete programs, the user would even be able to \textit{run} this minimal program, and work only with the simpler resulting trace.

\paragraph{User Study:} While these methods appear to have intuitive use in understanding type systems and type error debugging, the real-world effectiveness should be explored by user studies. A study on the methods effectiveness for the learning aspect, involving new users (e.g. students) would be feasible.

\bibliography{refs.bib}

\end{document}